\documentclass{moriond}

\def\Journal#1#2#3#4{{#1} {\bf #2}, #3 (#4)}

\def\PLB{{\em Phys. Lett.}  B}

\def\JHEP{\em JHEP}
\def\JINST{\em JINST}

\def\mgg{\ensuremath{m_{\gamma\gamma}}}
\def\Zee{\ensuremath{Z\to ee}}
\def\mllgg{\ensuremath{m_{\ell\ell\gamma\gamma}}}

\newcommand{\Photo}{\includegraphics[height=35mm]{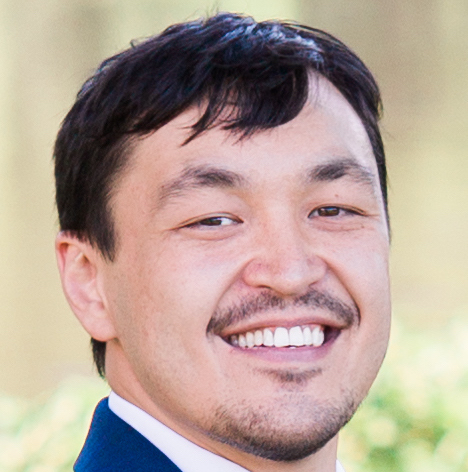}}

\begin{document}
\vspace*{4cm}
\title{Direct Searches at CMS}
\author{Joseph Reichert, on behalf of the CMS Collaboration}
\address{Laboratory for Elementary-Particle Physics, Cornell University,\\Ithaca, NY 14853, USA}

\maketitle\abstracts{Several of the CMS experiment's latest results on direct searches for new physics are presented. In particular, an emphasis is made to highlight the new models, unexplored final states, and innovative tools for discovery that these searches focus on.}

\section{Introduction}

One of the primary goals of the CMS experiment~\cite{CMS} is to perform searches for new phenomena, in pursuit of potential discoveries of beyond standard model (BSM) physics. This work details only a few of the $>$30 new BSM search results that CMS has released since last year's Moriond QCD conference, while other contributions to these proceedings cover most of the other new searches~\cite{SUSY,Resonances,LFV,DM,ExoHiggs,LLPs}. As many searches for BSM have already been performed at the LHC over the past 10\texttt{+} years, it is crucial that we continue to pursue innovative and novel ideas for discovering BSM physics, which has been an important focus for many of these searches. Note that all searches presented in this work are based on the Run~2 (2016--2018) dataset collected by CMS at $\sqrt{s} = 13$ TeV.

\section{Searches in photon-based final states}
\subsection{Search for an additional Higgs boson in the low mass diphoton final state}
This search~\cite{CMS-PAS-HIG-20-002} is designed to follow-up on an excess previously observed at $\mgg = 95$~GeV in the 2012 and 2016 CMS datasets~\cite{LowMassDiphoton2016}. The search relies on dedicated boosted decision trees (BDTs) to identify photons, select the primary interaction vertex, and to separate signal events from background. In particular, due to the proximity in mass to the $Z$ boson peak, the search has important backgrounds due to electrons that mimic the signature of a photon. These are suppressed via stringent electron vetoes and a requirement that the tracks associated to the primary interaction vertex be inconsistent with that typical of \Zee. The diphoton invariant mass distribution, which is used to extract signal, is shown in Figure~\ref{fig:diphoton} (left). The excess at $\mgg = 95$~GeV, while still present in the full Run~2 dataset with a global significance of $1.3\sigma$, did not grow with luminosity, as seen in Figure~\ref{fig:diphoton} (middle). Inclusive limits are set on the product of cross section and branching ratio of this additional Higgs boson decaying to photon pairs at 95\% confidence level (CL) at the level of 15--75 fb for masses in the range 70--110~GeV.

\subsection{Search for exotic Higgs decays to a Z boson and a light axion-like particle}
\begin{figure}[h]
\centering
\includegraphics[width=0.31\textwidth]{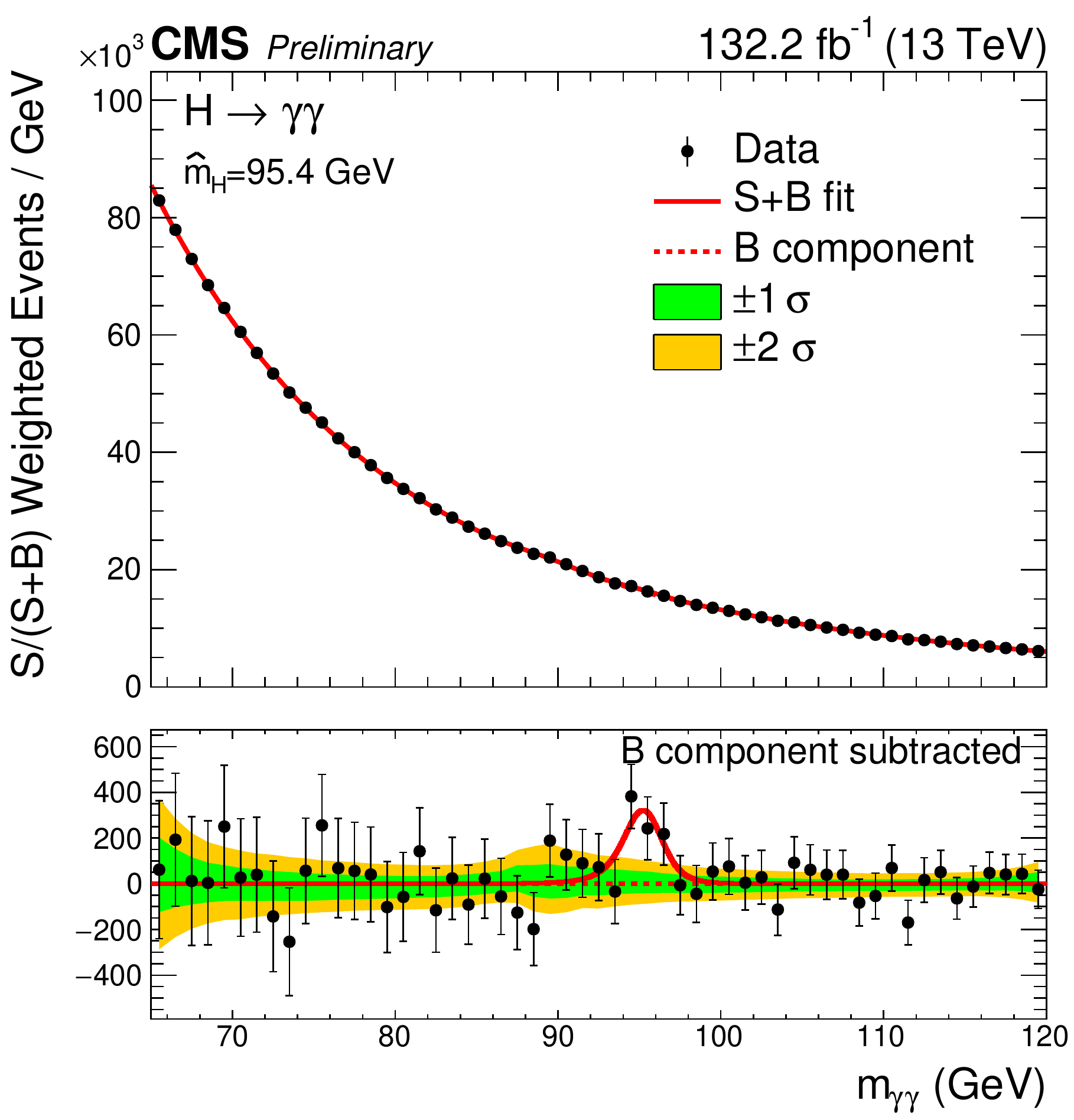}
\includegraphics[width=0.33\textwidth]{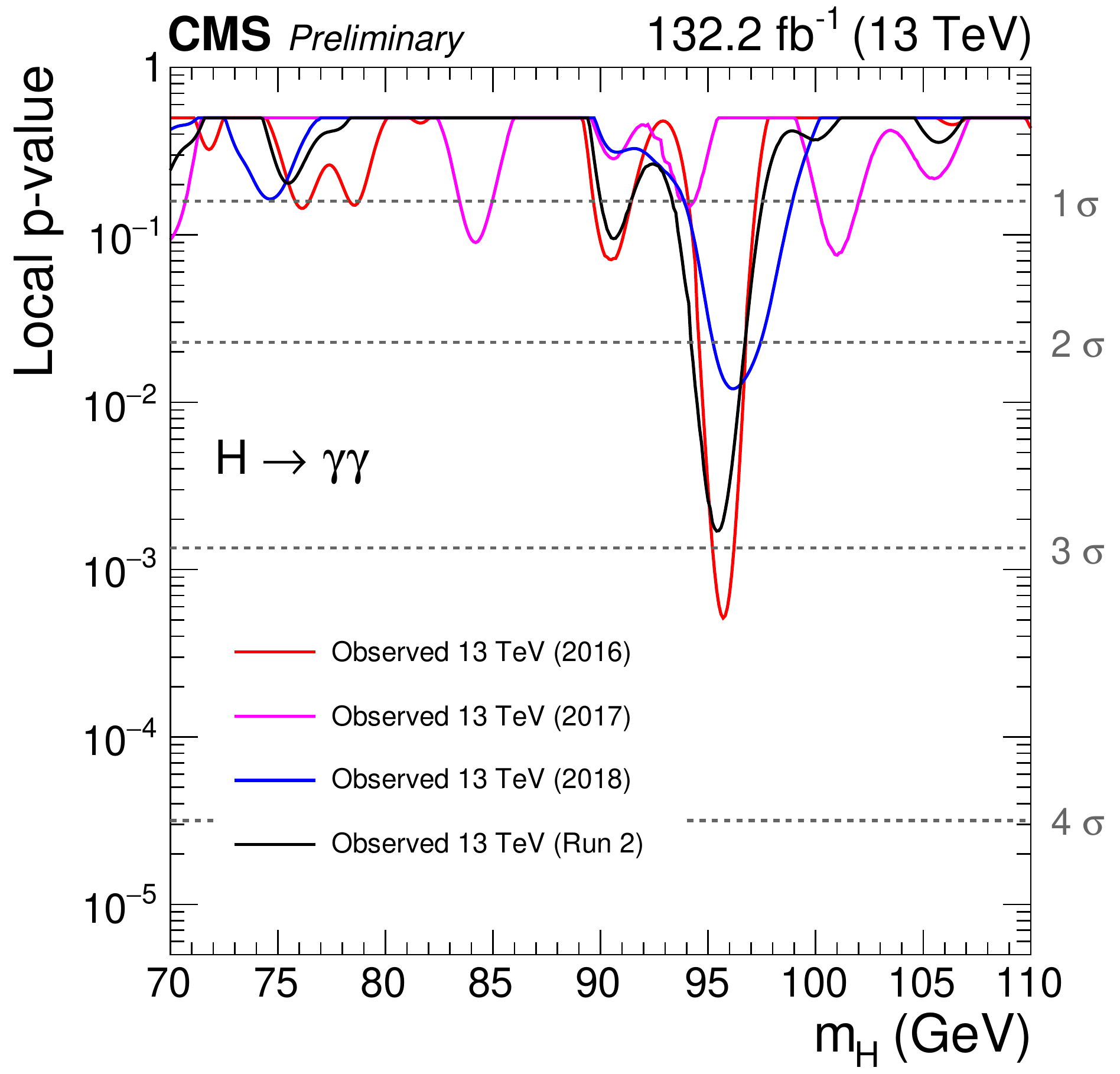}
\includegraphics[width=0.33\textwidth]{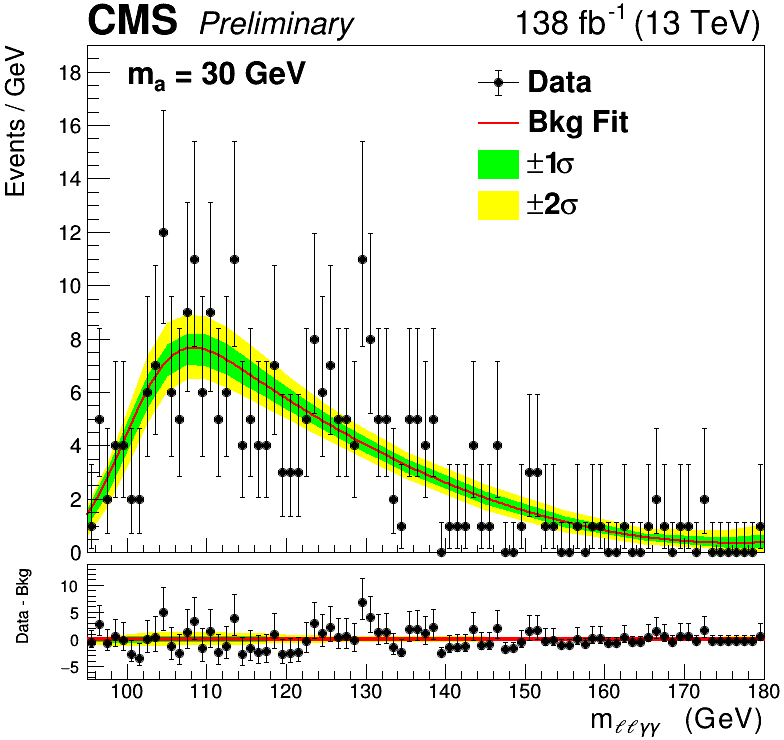}
  \caption{Low mass diphoton search~\protect\cite{CMS-PAS-HIG-20-002}: $\mgg$ distribution of selected events, along with a fit of the signal-plus-background model under a signal mass hypothesis of 95.4~GeV (left). The observed local $p$-values for an additional Higgs boson decaying to photon pairs, as a function of $m_H$, are shown for 2016, 2017, 2018, and their combination (middle); the most significant excess has a global significance of $1.3\sigma$. $H\to Za$ search~\protect\cite{CMS-PAS-HIG-22-003}: $\mllgg$ distribution of selected events (right), along with a fit of the background distribution based on a convolution of a falling spectrum background function with a step function and a Gaussian; these convolutions are necessary to model the turn-on of the $\mllgg$ distribution due to the 10~GeV photon transverse momentum requirement used in the search. The 115--135~GeV mass window is used to extract signal, while the remaining mass ranges are used as sidebands for determining the background. This is shown for $m_{a} = 30$~GeV, and a unique background fit and signal extraction is performed for each $m_{a}$ hypothesis considered.}
\label{fig:diphoton}
\label{fig:HZa}
\end{figure}

This first-of-its-kind search~\cite{CMS-PAS-HIG-22-003} for exotic Higgs decays to a $Z$ boson and a light axion-like particle ($a$, or ALP) is designed to probe ALPs decaying to two photons in the mass range 1--30~GeV. Due to the large boost of the ALP, the photons tend to overlap, which requires modifications to the photon identification criteria used in the search. The search also employs a BDT based on event kinematics, the mass hypothesis of interest, and photon identification quantities to reject background. The $\mllgg$ distribution is used to extract signal in the mass window 115--135~GeV, and background events are estimated via fits performed in the mass sidebands (95--115~GeV and 135--180~GeV), as seen in Figure~\ref{fig:HZa} (right). No significant excesses are observed, and limits are set at the 1--20 fb level on the product of cross section and branching ratio at 95\% CL.

\section{Searches for long-lived particles}
\subsection{Search for inelastic dark matter via displaced muons and missing transverse momentum}
This search~\cite{CMS-PAS-EXO-20-010} targets a model with two inelastically-coupled dark matter states that interact via a dark photon ($A'$). The mass splitting between the two dark matter states tends to be small, which means the heavier dark matter particle has a long lifetime before decaying to the stable, lighter dark matter state and an off-shell $A'$. The off-shell $A'$ decays to fermion pairs; here, the signature of interest is a pair of soft, displaced muons. As the standard muon reconstruction is inefficient at large production distances, a displaced reconstruction relying on tracks formed entirely in the muon chambers is also used in the search; this displaced algorithm provides an efficiency of roughly 95\%, compared with an efficiency for the standard algorithm of 60\% (0\%) for a 20~cm (1~m) displacement. Event kinematics, muon isolation criteria, and muon transverse impact parameters are used to suppress and estimate backgrounds. The search categorizes by the number of displaced muons that match in $\Delta R$ to a standard muon (0-, 1-, or 2-matches), where fewer matches is indicative of larger displacements. Across these three categories, a total of $2.2 \pm 0.7$ background events are predicted, which is consistent with the observation of 2 events in data. Limits are set on a wide range of dark matter masses, mass splittings, interaction strengths, and the product of cross section and branching ratio, as shown in Figure~\ref{fig:iDM} (left). This is the first search for inelastic dark matter at a hadron collider.

\subsection{Search for long-lived heavy neutral leptons using a displaced jet tagger}
This search~\cite{CMS-PAS-EXO-21-013} targets heavy neutral leptons ($N$) produced in $W$ boson decays, in scenarios with small $N$ masses or couplings leading to long lifetimes for $N$. The final state consists of a prompt lepton from the $W$ decay as well as a displaced lepton and a displaced jet; a dedicated deep neural network (DNN) is used to tag the displaced jet. The search categorizes events based on the lepton flavors and charges, the $\Delta R$ between the displaced lepton and displaced jet, and the displaced jet transverse impact parameter significance. Displaced objects are calibrated via standard model sources such as $\gamma \to ee$ and displaced $J/\psi \to \mu\mu$ decays, and backgrounds are estimated via sidebands of the tagger score and the invariant mass of the two leptons and the displaced jet. No significant excesses are seen, and limits are set at 95\% CL on the $N$ mass and the coupling strength between $N$ and $e$, $\mu$, or $\tau$ leptons, as seen in Figure~\ref{fig:HNL} (middle). Limits are set on both Dirac and Majorana heavy neutral leptons, as well as mixed coupling scenarios among the three lepton generations.

\section{Searches for heavy resonances}
\subsection{Search for W' bosons decaying to a top and a bottom quark in leptonic final states}
\begin{figure}[h]
\centering
  \includegraphics[width=0.335\textwidth]{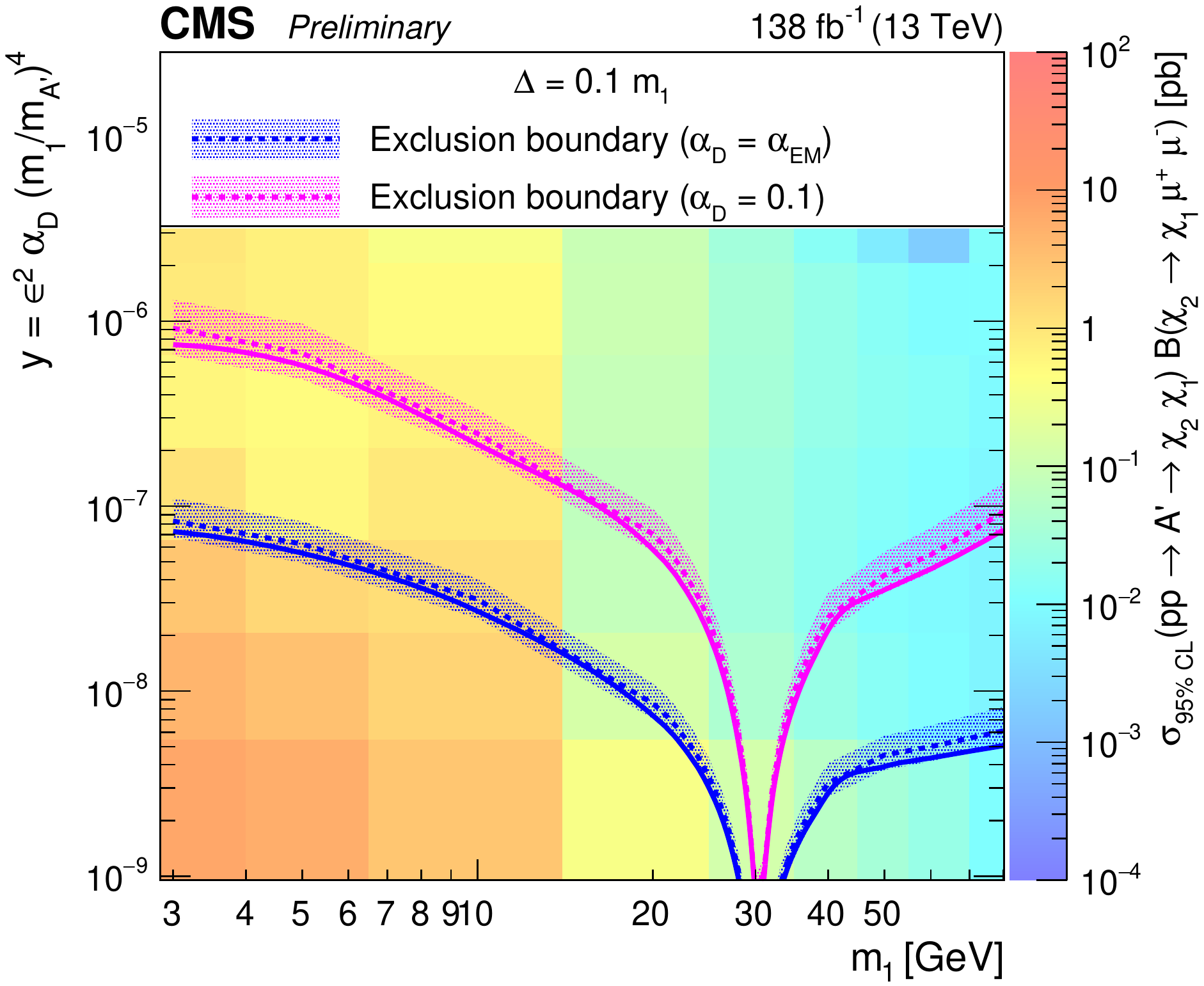}
  \includegraphics[width=0.31\textwidth]{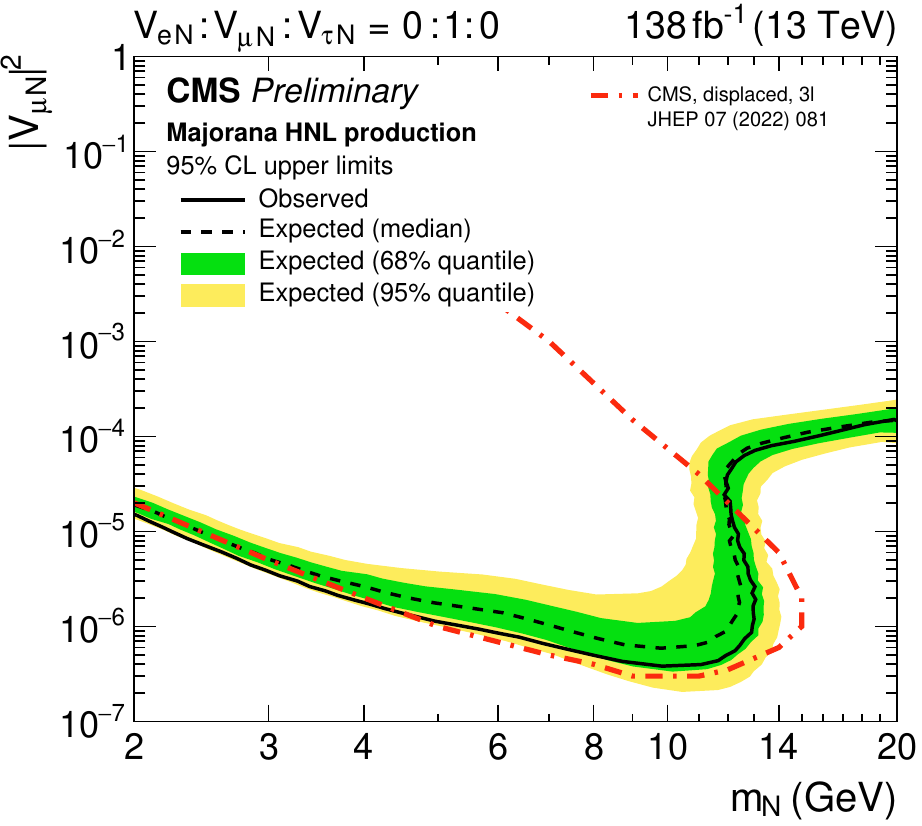}
  \includegraphics[width=0.34\textwidth]{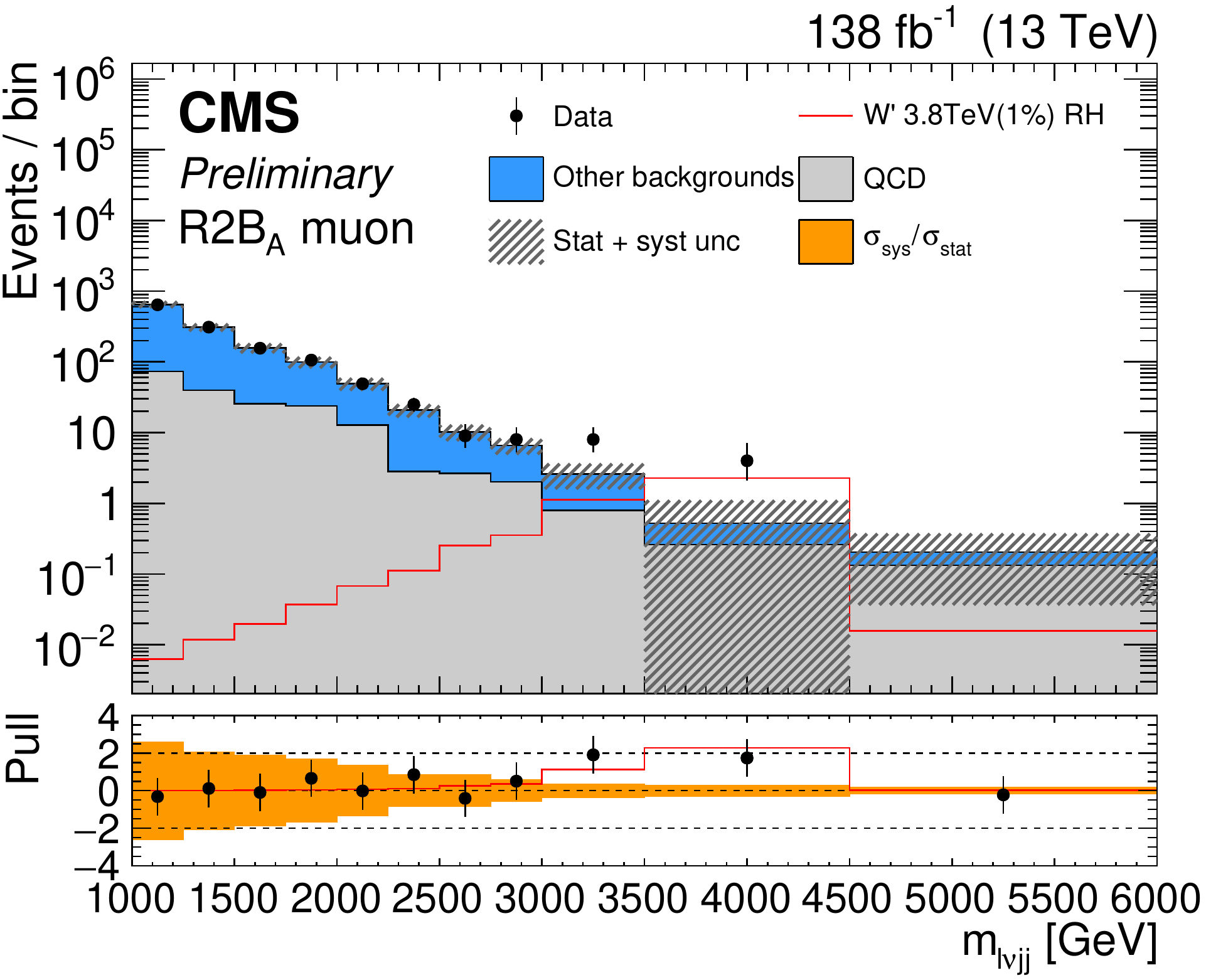}
  \caption{
Inelastic dark matter search~\protect\cite{CMS-PAS-EXO-20-010}: Limits on the lighter dark matter particle mass ($m_{1}$), interaction strength ($y$), and the product of dark matter cross section and branching ratio to muon pairs are shown for a given mass splitting $\Delta$ between the two dark matter states (left). Exclusions are shown for two hypotheses for the coupling parameter $\alpha_{D}$. Heavy neutral lepton search~\protect\cite{CMS-PAS-EXO-21-013}: Expected and observed 95\% CL limits on Majorana heavy neutral lepton production (middle), shown here for pure muon couplings, as a function of $N$ mass and coupling strength. A complementary CMS search for heavy neutral leptons in the trilepton final state is also overlaid~\protect\cite{HNL3L}.
$W' \to tb$ search~\protect\cite{CMS-PAS-B2G-20-012}: $m_{\ell\ell\nu j}$ distribution for selected events in the signal region with two $b$-jets and a muon, after the signal-plus-background fit (right).}
\label{fig:iDM}
\label{fig:HNL}
\label{fig:Wprime}
\end{figure}

This search~\cite{CMS-PAS-B2G-20-012} for $W' \to tb$, with $t \to \ell\nu b$, uses the known $W$ boson mass to constrain the neutrino momentum. This, along with several other criteria, is used to reconstruct the top quark kinematics and subsequently fully reconstruct the $W'$ mass. Due to the two $b$ quarks expected in the final state, events with no $b$-jets are used as a background-dominated control region, while the signal regions are binned based on the $b$-jet pairings with the top quark or the $W'$. Backgrounds are estimated via mass sidebands in data. Figure~\ref{fig:Wprime} (right) shows the $m_{\ell\ell\nu j}$ distribution in the signal region with two $b$-jets and a muon; overall, a $2.6\sigma$ local ($2.0\sigma$ global) excess is seen at 3.8 TeV for a narrow $W'$ with right-handed chirality. Limits are set at 95\% CL on a range of masses, chiralities, and widths (up to 30\%, for the first time), and the search provides the most stringent limits on $W' \to tb$ to date.

\subsection{Search for Higgs boson pair production in the bbWW final state}
This search~\cite{CMS-PAS-HIG-21-005} for resonant Higgs boson pair production in the $X\to HH\to bbWW$ final state benefits from the large Higgs branching ratio to $b$ quarks and $W$ bosons, as well as the high lepton trigger efficiency, but has large $t\bar{t}$ backgrounds to contend with. A set of mass-parameterized DNNs are used with a Lorentz Boost Network as a pre-processor to compute useful input variables. The DNN score distributions are used to extract signal, as well as a heavy mass estimator in the dilepton channel which determines the most likely mass of the $X$ resonance, after accounting for neutrinos. For a spin-0 resonance, limits are set on the $X\to HH$ cross section at 95\% CL for $m_{X}$ in the range 250--900~GeV at a level of 20 fb to 7 pb. The search is additionally performed for spin-2 resonances and for non-resonant production of Higgs boson pairs.

\section{Summary}

This work reports on several searches for BSM physics performed at CMS using the full Run~2 dataset. Stringent limits are set by CMS on a number of models and final states~\cite{EXOSummary,B2GSummary}; however, there are many reasons to be optimistic about the future. Several mild excesses have been seen, which will benefit from larger datasets during Run~3 and the HL-LHC to distinguish statistical fluctuations from new physics. Furthermore, the CMS Collaboration is continuing to develop novel ideas and innovations in pursuit of a BSM discovery.

\section*{Acknowledgments}

The author would like to thank the CMS Collaboration for the opportunity to present, and the organizers of the $57^\mathrm{th}$ Rencontres de Moriond for the successful conference. This work is supported by the U.S. National Science Foundation under award number NSF-PHY-2209443.

\end{document}